\documentclass[11 pt]{amsart}
\newtheorem{theorem}{Theorem}[section]
\newtheorem{lemma}[theorem]{Lemma}
\newtheorem{proposition}[theorem]{Proposition}

\theoremstyle{definition}

\newtheorem{remark}[theorem]{Remark}

\numberwithin{equation}{section}

\DeclareMathOperator*{\esssup}{ess\,sup}

\theoremstyle{remark}
\usepackage{amscd,amssymb,color}

\begin{document}
\baselineskip 17 pt

\title[]
{Asian Option Pricing under Uncertain Volatility Model}

\author[Yue-Cai Han and Chun-Yang Liu ]
{Yue-Cai Han and Chun-Yang Liu}



\begin{abstract}
In this paper, we study the asymptotic behavior of Asian option prices in
the worst case scenario under an uncertain volatility model. We give a procedure to approximate the Asian option prices with a small volatility interval. By
imposing additional conditions on the boundary condition and cutting the obtained Black-Scholes-Barenblatt equation into two Black-Scholes-like equations,
we obtain an approximation method to solve the fully nonlinear PDE.
\end{abstract}

\maketitle
\section{Introduction}
An option on a traded account is a financial contract which allows the buyer of the
contract obtains the right to trade an underlying asset for a specified price, called strike price,
during the life of the option. There are varieties of options, such as European option, American option, Asian option and barrier option.
As the foundations for the modern analysis of options, the Black-Scholes-Merton pricing formula for European option was introduced by Black, Scholes \cite{BS} and Merton \cite{M}.
In Blcak-Scholes-Merton model, the volatility is assumed to be constant.
However, constant volatility cannot explain observed market prices for options.
\par
After Black, Scholes and Merton's work, some scholars studied option pricing models with stochastic volatility. In a series of papers, several models for stochastic volatility were introduced, such as Hull-White stochastic volatility model \cite{HW} and Heston stochastic volatility model \cite{H}.
\par Uncertain volatility model is another approach to describe the non-constant volatility.
In 1995, uncertain volatility model was introduced by Lyons \cite{Lyons} and Avellaneda et al. \cite{Avellaneda}. In these models, volatility is assumed to lie within a range of values. So the prices are no longer unique. We can only get the best-case scenario prices and the worst-case scenario prices. Several problems about uncertain volatility have been studied. We can see these results in Lyons \cite{Lyons}, Avellaneda et al. \cite{Avellaneda}, Dokuchaev, Savkin \cite{DS}, Forsyth, Vetzal \cite{FV} and Vorbrink \cite{JV}. Pricing in uncertain volatility models involving nonlinear partial differential equations have been showed in their paper. Some numerical methods have been proposed in Pooley, Forsyth, Vetzal\cite{DPK}, and
Avellaneda et al. \cite{Avellaneda}.
\par In 2014, Fouque and Ren \cite{Jean} studied the price of European derivatives in the worst case scenario under the uncertain volatility model. They provided an approximate method of pricing the derivatives with a small volatility interval. In addition, the paper also presented that the solution reduces to a constant volatility problem when it comes to simple options with convex payoffs.
\par
In this paper, the pricing problem of Asian options is studied. The payoff function is path-dependent on risky asset price processes. Another variable is given to solve the problem. In the process of finding the estimation of the worst case scenario Asian option prices, the first problem that we meet is obtaining the Hamilton-Jacobi-Bellman (HJB) equation of the prices. The HJB equation is called Black-Scholes-Barenblatt (BSB) equation in the financial mathematics. We can get the BSB equation by the stochastic control theory. The next difficulty is to proof the convergence of the estimation. To control the error term, we obtain its expectation form by Dynkin formula and  find what conditions should we impose on the payoff function by proving and deducing. Finally, we get the approximation procedure for the prices. Compare with Fouque and Ren's paper \cite{Jean}, we add an equation in the stochastic control system and it can also be reflected in the BSB equation. In terms of the dynamic of the risky asset price process, we give an equation to describe the path-dependence. When estimate the expectation form, We use the relationship between the two processes. In section 4.4, we fix one of two variables first to simplify the problem. Another method we used to manage the two variables is changing the form of the BSB equation.
\par
The organization of this paper is as follows. In section 2, we briefly describe Asian options under uncertain volatility model and give the Black-Scholes-Barenblatt (BSB) equations of option prices. In section 3, we find the estimation of Asian option prices in the worst case scenario and the estimation is relied on two Black-Scholes-like PDEs. Next, we propose the main result of this paper which shows the rationality of estimation. In section 4, we give the proof of
the main result. Through the conditions imposed on the payoff function, we get the convergence of the error term. In the process, we obtain the expectation form of the error term and it is cut into three parts. The controls of the three parts are given by the stochastic control theory and the properties of the worst-case scenario Asian option price process. Finally, we give the conclusion of this paper.

\section{Asian options under uncertain volatility model}

 In this section, we introduce the Asian options under uncertain volatility model. Then we give the Black-Scholes-Barenblatt (BSB) equation of the Asian options' prices. Suppose that $\mathcal{X}$ is an Asian option written on the risky asset with maturity T and payoff
$\varphi(\cdot)$.
 $\varphi(\cdot)$ is a non-convex function and the result is identical to Black-Scholes result under convex condition. That is to say, the results of this article cover generalized Asian options. 
 \par Assume that the price process of the risky asset $X_t$ solves the following stochastic differential equation
\begin{equation}\label{e2.1}
dX_t=rX_tdt+\sigma_tX_tdW_t,
\end{equation}
where $r$ is the constant risk-free interest rate, $W_t$ is a standard Brownian motion on the probability space $(\Omega,\mathfrak{F},\mathbb{P})$ and the volatility process $\sigma_t \in\mathcal{A}[\underline{\sigma},\overline{\sigma}]$ for each $t\in [0,T]$, which is a family of progressively measurable and $[\underline{\sigma},\overline{\sigma}]$-valued processes.
By the definition as above, we know that the volatility in an uncertain volatility model is not a stochastic process with
a probability distribution, but a family of stochastic processes with unknown prior information. Thus, what can we use to
distinguish the difference between uncertain volatility models is the model ambiguity.

\par Due to the path-dependence on risky asset price processes, we assume that $Y_{t,T}$ satisfies the expression as follows
\begin{equation}\label{e2.2}
Y_{t,T}=\frac{Y_T-Y_t}{T-t},
\end{equation}
where $Y_t=\int^t_0 X_u du$.
\par Then we can get Asian option prices in the worst case scenario price at time $t<T$ as follows 
\begin{equation}\label{e2.3}
V(t,X_t,Y_t)=\exp{(-r(T-t))}~\underset{\sigma\in\mathcal{A}[\underline{\sigma},\overline{\sigma}]}{\esssup}~ E[\varphi(Y_{0,T})|\mathfrak{F}_t],
\end{equation}
where $\esssup$ is essential supremum. By the ambiguity of the uncertain volatility model, we obtain the definition of price as
equation (\ref{e2.3}). Obviously, the
worst case scenario price is for the seller of options.
It is related to coherent risk measure which quantifies the model risk
induced by volatility uncertainty 
(see \cite{R}). Moreover, the model ambiguity in mathematical finance has captured the attention of many. Therefore, we should pay attention to the importance of the worst case prices.

\par Through the stochastic control theory (see  \cite{YZ}),
$V(t,X_t,Y_t)$ satisfies the Hamilton-Jacobi-Bellman (HJB) equation (Black-Scholes-Barenblatt (BSB) equation).
\begin{lemma}     \label{BSB}
$V(t,X_t,Y_t)$ satisfies the following Black-Scholes-Barenblatt equation
\begin{eqnarray}\label{e2.4}
\qquad\left\{ \begin{array}{rcl}
\partial_tV+r(x\partial_x V-V)+x\partial_y V +\sup\limits_{\sigma\in\mathcal{A}[\underline{\sigma},\overline{\sigma}]}\left[\frac{1}{2}
x^2\sigma^2\partial^2_{xx}V\right]=0, \\
0\leq t\leq T,~x\geq 0,~y\geq 0,\\
V(T,x,y)=\varphi(\frac{y}{T}),~x\geq 0,~y\geq 0.\\
\end{array}
\right.
\end{eqnarray}
\end{lemma}
{\em Proof}.
 Notice that the stochastic control system is
\begin{equation*}
\qquad\left\{ \begin{array}{rcl}
d X_t&=&r X_t dt+\sigma_t X_t d W_t ,\qquad \sigma_t\in\mathcal{A}[\underline{\sigma},\overline{\sigma}],\\
d Y_t&=&X_t dt.
\end{array}
\right.
\end{equation*}
Then for all $(s,x,y)\in[0,T]\times R^+\times R^+$, we first establish the dynamic program frame
\begin{eqnarray}\label{e2.5}
\qquad\left\{ \begin{array}{rcl}
d X_t&=&r X_t dt+\sigma_t X_t d W_t , \\
d Y_t&=&X_t dt,\\
X_s&=&x,\\
Y_s&=&y.
\end{array}
\right.
\end{eqnarray}
The cost function is
$$J(s,x,y;\sigma)=E_s\left[e^{-r(T-s)}\varphi(Y_{0,T})\right],$$
where $E_s[\cdot]=E[\cdot|\mathfrak{F}_s]$. The value function is
\begin{eqnarray*}
V(s,x,y)&=&\underset{\sigma\in\mathcal{A}[\underline{\sigma},\overline{\sigma}]}{\esssup}~J(s,x,y;\sigma).
\end{eqnarray*}

\par For all $0\leq s\leq \hat{s}\leq T$, $\sigma\in\mathcal{A}[\underline{\sigma},\overline{\sigma}]$, we have
\begin{eqnarray*}
V(s,x,y)&\geq& E_s\left[e^{-r(T-s)}\varphi(Y_{0,T})\right]\\
&=&E_s\left[\int^{\hat{s}}_{s}-re^{-r(T-t)}\varphi dt+e^{-r(T-\hat{s})}\varphi\right].
\end{eqnarray*}
Then we obtain
$$0\geq E_s\left[\int^{\hat{s}}_{s}-re^{-r(T-t)}\varphi dt\right]+V(\hat{s},x,y)-V(s,x,y).$$
Divided by $\widehat{s}-s$ on both sides of the inequality, we have that
$$0\geq E_s\left[\frac{\int^{\hat{s}}_{s}-re^{-r(T-t)}\varphi dt}{\hat{s}-s}\right]+
\frac{V(\hat{s},x,y)-V(s,x,y)}{\hat{s}-s}.$$
\par Here, assume that $\varphi$ is Lipschitz continuous. Then according to It$\hat{\rm o}$ formula and equations (\ref{e2.5}), we obtain
\begin{eqnarray*}
dV&=&V_t dt+V_x d X_t+V_y d Y_t+\frac{1}{2}V_{xx}d X_t d X_t+\frac{1}{2}V_{yy}d Y_t d Y_t+\frac{1}{2}V_{xy}d X_t d Y_t\\
&=&(V_t+rX_t V_x+X_t V_y+\frac{1}{2}\sigma^2_t X_t^2 V_{xx})dt+\sigma_t X_t V_x d W_t.
\end{eqnarray*}
Let $\hat{s}\rightarrow s$. For all $\sigma\in\mathcal{A}[\underline{\sigma},\overline{\sigma}]$, we have that
\begin{eqnarray*}
0&\geq&-r E_s[e^{-r(T-s)}\varphi]+V_t+r X_s V_x+X_s V_y+\frac{1}{2}\sigma^2_s X_s^2V_{xx}\\
&\geq&-r V(s,x,y)+V_t(s,x,y)+rx V_x(s,x,y)+x V_y(s,x,y)+\frac{1}{2}\sigma^2_s X_s^2V_{xx}(s,x,y),
\end{eqnarray*}
which is
\begin{equation}\label{e2.6}
0\geq-rV+V_t+rxV_x+xV_y+\sup\limits_{\sigma\in\mathcal{A}[\underline{\sigma},\overline{\sigma}]}\frac{1}{2}\sigma^2x^2V_{xx}.
\end{equation}
\par On the other hand, for any $\varepsilon>0$, there is $\sigma(\varepsilon)\in\mathcal{A}[\underline{\sigma},\overline{\sigma}]$ such that
\begin{eqnarray*}
V(s,x,y)-\varepsilon(\hat{s}-s)&\leq &E_s\left[e^{-r(T-s)}\varphi\right]\\
&=&E_s\left[\int^{\hat{s}}_{s}-re^{-r(T-t)}\varphi dt\right]+E_s\left[e^{-r(T-\hat{s})}\varphi\right].
\end{eqnarray*}
So we have that
\begin{eqnarray*}
-\varepsilon&\leq&E_s\left[\frac{\int^{\hat{s}}_{s}-re^{-r(T-t)}\varphi dt}{\hat{s}-s}\right]+\frac{V(\hat{s},x,y)-V(s,x,y)}{\hat{s}-s}.
\end{eqnarray*}
Argument as above, we obtain
\begin{equation}\label{e2.7}
0\leq-rV+V_t+rxV_x+xV_y+\sup\limits_{\sigma\in\mathcal{A}[\underline{\sigma},\overline{\sigma}]}\frac{1}{2}\sigma^2x^2V_{xx}.
\end{equation}
Combining (\ref{e2.6}) with (\ref{e2.7}), we have
$$0=-rV+V_t+rxV_x+xV_y+\sup\limits_{\sigma\in\mathcal{A}[\underline{\sigma},\overline{\sigma}]}\frac{1}{2}\sigma^2x^2V_{xx}.$$
\hfill $\Box$
\begin{remark}
Here, adding variable $Y$ into dynamical system leads to a more complex stochastic control system, which adds the dimensionality of the BSB equation.
\end{remark}
\begin{remark}
Notice that, (\ref{e2.4}) is a fully nonlinear PDE which doesn't have a solution like Black-Scholes equation. Thus, we decide to solve the problem by reducing it to solving two Black-Scholes-like PDEs.
\end{remark}
\section{Black-Scholes-like PDEs and Main Result}\label{sec3}
 In this section, we first reparameterize the uncertain volatility model to study the prices in the worst case scenario. Assume that the risky asset price process $X_t^{\varepsilon}$ has a dynamic
\begin{eqnarray}\label{e2.8}
\qquad\left\{ \begin{array}{rcl}
d X_t^{\varepsilon}&=&r X_t^{\varepsilon} dt+\sigma_t X_t^{\varepsilon} d W_t , \\
d Y_t^{\varepsilon}&=&X_t^{\varepsilon} dt,
\end{array}
\right.
\end{eqnarray}
where $\sigma_t\in\mathcal{A}^{\varepsilon}=$\{$\sigma_t|\sigma_t$ is a $[\sigma_0,\sigma_0+\varepsilon]-$valued
processively measurable process\} and $\sigma_0\in[\underline{\sigma},\overline{\sigma}]$.
\par The cost function is
$$J^{\varepsilon}(t,x,y;\sigma)=e^{-r(T-t)}E_{txy}\left[\varphi(Y_{0,T}^{\varepsilon})\right],$$
where $E_{txy}[\cdot]$ means the conditional expectation taken with respect to $X_t^{\varepsilon}=x$, $Y_t^{\varepsilon}=y$.
The value function is
$$V^{\varepsilon}(t,x,y;\sigma)=~\underset{\sigma\in\mathcal{A}^{\varepsilon}}{\esssup}~[J^{\varepsilon}(t,x,y;\sigma)].$$
\par By Lemma \ref{BSB} we get the following Black-Scholes-Barenblatt equation of $V^{\varepsilon}$.
\begin{eqnarray}\label{e2.9}
\qquad\left\{ \begin{array}{rcl}
\partial_tV^{\varepsilon}+r(x\partial_xV^{\varepsilon}-V^{\varepsilon})+x\partial_yV^{\varepsilon}+ \sup\limits_{\sigma\in\mathcal{A}^{\varepsilon}}\frac{1}{2}x^2\sigma^2\partial^2_{xx}V^{\varepsilon}=0,\\
0\leq t\leq T,~x\geq 0,~y\geq 0,
\\V^{\varepsilon}(T,x,y)=\varphi(\frac{y}{T}),
~x\geq 0,~y\geq 0,
\end{array}
\right.
\end{eqnarray}
which is equivalent to
\begin{eqnarray}\label{e2.10}
\qquad\left\{ \begin{array}{rcl}
\partial_tV^{\varepsilon}+r(x\partial_xV^{\varepsilon}-V^{\varepsilon})+x\partial_yV^{\varepsilon}
+ \sup\limits_{\gamma\in\mathcal{A}[0,1]}\frac{1}{2}x^2(\sigma_0+\varepsilon\gamma)^2\partial^2_{xx}V^{\varepsilon}=0,
\\
0\leq t\leq T,~x\geq 0,~y\geq 0,
\\V^{\varepsilon}(T,x,y)=\varphi(\frac{y}{T}),
~x\geq 0,~y\geq 0,
\end{array}
\right.
\end{eqnarray}
where $\mathcal{A}[0,1]=$\{$\gamma_t|\gamma_t$ is a $[0,1]-$valued processively measurable process\}.
\par It is obvious that the worst case scenario price is larger than any Black-Scholes price with a constant volatility $\sigma_0\in[\underline{\sigma},\overline{\sigma}]$. We will show that the worst case scenario price of Asian option converges to its Black-Scholes price with constant volatility $\sigma_0$  in following section. In addition, the rate of convergence of the Asian option prices as the volatility interval shrinks to a single point can be obtained. Then we can get the estimation of the prices through this result when the interval is sufficiently small.
\par
Let $V_0$ be the Black-Scholes prices, $V^0=V^{\varepsilon}|_{\varepsilon=0}$,
$V_1=\partial_{\varepsilon}V^{\varepsilon}|_{\varepsilon=0}$. Now, we suppose that $V^{\varepsilon}$ is continuous with respect to $\varepsilon$. Then, by the continuity of $V^{\varepsilon}$ and equation (\ref{e2.3}), we have $V_0=V^0=V^{\varepsilon}|_{\varepsilon=0}$.
It's well known that $V_0$ satisfies the following partial differential equation.
\begin{eqnarray}\label{e2.11}
\qquad\left\{ \begin{array}{rcl}
\partial_t V_0+r(x\partial_x V_0-V_0)+x\partial_y V_0+\frac{1}{2}\sigma_0^2x^2\partial^2_{xx}V_0=0,\\
0\leq t\leq T,~x\geq 0,~y\geq 0,\\
V_0(T,x,y)=\varphi(\frac{y}{T}),~x\geq 0,~y\geq 0.
\end{array}
\right.
\end{eqnarray}
\par On the other hand, we have $V_1=\partial_{\varepsilon}V^{\varepsilon}|_{\varepsilon=0}$, which is the rate of convergence of the Asian option prices as $\varepsilon$ approaches 0. To obtain the equation characterizing $V_1$, we differentiate both sides of equations (\ref{e2.10}) with respect to $\varepsilon$ and let $\varepsilon=0$, then we have that
\begin{eqnarray}\label{e2.12}
\left\{ \begin{array}{r}
\partial_t V_1+r(x\partial_x V_1-V_1)+x\partial_y V_1+\frac{1}{2}\sigma_0^2x^2\partial^2_{xx}V_1
+\sup\limits_{\gamma\in\mathcal{A}[0,1]}\gamma\sigma_0x^2\partial^2_{xx}V_0=0,\\
0\leq t\leq T,~x\geq 0,~y\geq 0,\\
V_1(T,x,y)=0,~x\geq 0,~y\geq 0.
\end{array}\negthinspace\negthinspace\negthinspace\negthinspace\negthinspace\negthinspace\negthinspace
\right.
\end{eqnarray}
\par
Now, we have two Black-Scholes-like PDEs as above. Next, we want to find the connection between $V^{\varepsilon}$ and $V_0,~V_1$. Then we try to prove if we can impose additional conditions on the payoff function to make the error term $V^{\varepsilon}-(V_0+\varepsilon V_1)$ be of order $\circ(\varepsilon)$. That is to say, the estimation
of the worst case scenario Asian option prices will approach the truth-value as the model ambiguity vanishes. It will also show us a method to estimate the worst case Asian option prices. By the deducing in the next section, we get following theorem which is the main result of this paper.
\begin{theorem}     \label{main}
Assume that $\varphi\in\mathcal{C}_p^2(R^+)$ is Lipschitz continuous, and the second derivative of $\varphi$
is continuous.
Then
\begin{equation*}  \label{e2.13}
\lim_{\varepsilon\downarrow0}\frac{V^{\varepsilon}-(V_0+\varepsilon V_1)}{\varepsilon}=0.
\end{equation*}
\end{theorem}
Here $\varphi\in\mathcal{C}_p^2(R^+)$ means that 
its derivatives up to order 2 have polynomial growth.
\begin{remark}
To prove the theorem \ref{main}, there are some difficulties. The first one is how to convert the error term into an estimable form. We get its expectation form and cut it into three parts in next section. The second difficulty is how to estimate the three parts. We will use the stochastic control theory, the zero set property of the equation (\ref{e3.3}), the properties of the sublinear expectation\cite{LMS} and the properties of the worst case scenario Asian option price processes.
\end{remark}
\begin{remark}
By theorem \ref{main}, we can compute Asian option price $V^\varepsilon(t,X_{t}^{\varepsilon},Y_{t}^{\varepsilon})$ with its approximation, $V_0(t,X_{t}^{\varepsilon},Y_{t}^{\varepsilon})+\varepsilon V_1(t,X_{t}^{\varepsilon},Y_{t}^{\varepsilon})$, where $V_0(t,X_{t}^{\varepsilon},Y_{t}^{\varepsilon})$ is the  Black-Scholes price of Asian option and $V_1(t,X_{t}^{\varepsilon},Y_{t}^{\varepsilon})$ can be numerically computed by a simple difference scheme according to (\ref{e2.12}).(see \cite{DPK})
\end{remark}
\begin{remark}
Notice that (\ref{e2.11}) and (\ref{e2.12}) are independent of $\varepsilon$.  So when we compute $V^\varepsilon$ with different $\varepsilon$, we just need to compute $V_0$ and $V_1$ once for all small values of $\varepsilon$ by Theorem \ref{main}.
\end{remark}

\section{The proof of the main result}
In this section, we try to control the error term to prove that we can compute $V^{\varepsilon}$ with its estimation $V_0+\varepsilon V_1$.
As the conditions imposed on $\varphi$ which mentioned in Theorem \ref{main}, we have following process of proof. On the other hand, our thinking process is also reflected in the next parts.
\subsection{The Lipschitz continuity of payoff function}
From section 3 we know that only with the continuity of $V^{\varepsilon}$ can we obtain the PDEs of $V_0$ ($=V^{\varepsilon}|_{\varepsilon=0}$) and $V_1$ ($=\partial_{\varepsilon}V^{\varepsilon}|_{\varepsilon=0}$). Thus, to get the continuity of $V^{\varepsilon}$, we suppose that $\varphi$ is Lipschitz continuous.
Then, there exists a constant $K_1$ such that
$$|\varphi(x)-\varphi(y)|\leq K_1|x-y|,~for~all~x\neq y,~x,y\in R^+.$$
Thus, we have Lemma as follows.
\begin{lemma}  \label{L}
Assume that $\varphi$ is Lipschitz continuous. Then $V^{\varepsilon}$ is continuous with respect to $\varepsilon$.
\end{lemma}
{\em Proof}.
 Let $0\leq\varepsilon_0\leq\varepsilon<1$. Notice that
$$V^{\varepsilon}(t,x,y;\sigma)=~\underset{\sigma\in\mathcal{A}^{\varepsilon}}{\esssup}~
\left\{e^{-r(T-t)}E_{txy}\left[\varphi(Y_{0,T}^{\varepsilon})\right]\right\}.$$
We have that
\begin{eqnarray*}
e^{r(T-t)}V^{\varepsilon_0}(t,x,y;\sigma)&=&~\underset{\sigma\in\mathcal{A}^{\varepsilon_0}}{\esssup}~
E_{txy}\left[\varphi(Y_{0,T}^{\varepsilon_0}(\sigma))\right]\\
&=&~\underset{\sigma\in\mathcal{A}^{\varepsilon}}{\esssup}~
E_{txy}\left[\varphi(Y^{\varepsilon}_{0,T}(\sigma\wedge(\sigma_0+\varepsilon_0)))\right].
\end{eqnarray*}
By the Lipschitz continuity of $\varphi$ and equation (\ref{e2.1}), there is a constant $K_1$ such that
\begin{eqnarray*}
&&e^{r(T-t)}\left|V^{\varepsilon}(t,x,y;\sigma)-V^{\varepsilon_0}(t,x,y;\sigma)\right|\\
&\leq&~\underset{\sigma\in\mathcal{A}^{\varepsilon}}{\esssup}~\left|E_{txy}\left[\varphi\left(Y_{0,T}^{\varepsilon}(\sigma)\right)\right]
-E_{txy}\left[\varphi\left(Y^{\varepsilon}_{0,T}(\sigma\wedge(\sigma_0+\varepsilon_0))\right)\right]\right|\\
&\leq&K_1~\underset{\sigma\in\mathcal{A}^{\varepsilon}}{\esssup}~\left(E_{txy}\left|Y_{0,T}^{\varepsilon}(\sigma)-
Y^{\varepsilon}_{0,T}(\sigma\wedge(\sigma_0+\varepsilon_0))\right|^2\right)^{1/2}\\
&\leq&\left(K_1/T\right)~\underset{\sigma\in\mathcal{A}^{\varepsilon}}{\esssup}~\left(E_{txy}\int^T_0
\left|X_u^{\varepsilon}(\sigma)-X_u^{\varepsilon}(\sigma\wedge(\sigma_0+\varepsilon_0))\right|^2du\right)^{1/2}.
\end{eqnarray*}
With the estimates of the moments of solutions of stochastic differential equations (Theorem 9 in Section 2.9 and Corollary
12 in section 2.5 of 
\cite{NV}),
there are constants $N=N(q,r,\sigma_0),~N'=N'(q,r,\sigma_0),~and~C=\max\{NN',~N+N'\}$ such that
\begin{eqnarray*}
&E_{txy}&\left[\sup\limits_{s\leq u}\left|X_s^{\varepsilon}(\sigma)-X_s^{\varepsilon}(\sigma\wedge(\sigma_0+\varepsilon_0))\right|^{2q}\right]\\
&\leq&
Nu^{q-1}e^{Nu}E_{txy}\left[\int^u_0\left|X_s^{\varepsilon}(\sigma)\right|^{2q}\cdot\left|\sigma_s-\sigma_s\wedge(\sigma_s+\varepsilon_0)\right|^{2q}ds\right]\\
&\leq&Nu^{q-1}e^{Nu}N'e^{N'u}u(1+x^{2q})|\varepsilon-\varepsilon_0|^{2q}\\
&=&Cu^qe^{Cu}(1+x^{2q})|\varepsilon-\varepsilon_0|^{2q}.
\end{eqnarray*}
Thus we have that
\begin{eqnarray*}
&&e^{r(T-t)}|V^{\varepsilon}(t,x,y)-V^{\varepsilon_0}(t,x,y)|\\
&\leq& \left(K_1/T\right)~\underset{\sigma\in\mathcal{A}^{\varepsilon}}{\esssup}~\left(\int_0^{T}E_{txy}\sup\limits_{s\in[0,u]}
|X_u^{\varepsilon}(\sigma)-X_u^{\varepsilon}(\sigma\wedge(\sigma_0+\varepsilon_0))|^2du\right)^{1/2}\\
&\leq&\left(K_1/T\right)~\underset{\sigma\in\mathcal{A}^{\varepsilon}}{\esssup}~
\left(\int^T_0 Cue^{Cu}(1+x^2)|\varepsilon-\varepsilon_0|^2du\right)^{1/2}\\
&\leq&K'_1(1+x^2)^{1/2}|\varepsilon-\varepsilon_0|,
\end{eqnarray*}
where $K'_1=K'_1(K_1,C,T)$.
\par Let $\varepsilon\rightarrow\varepsilon_0$. We have that $|V^{\varepsilon}(t,x,y)-V^{\varepsilon_0}(t,x,y)|\rightarrow0.$
\par The continuity of $V^\varepsilon$ with respect to $\varepsilon$ can be proved similarly when $\varepsilon\leq\varepsilon_0$.
\par\qquad
\hfill $\Box$

\subsection{Expectation form of the error term}
In this part, we analyze the error term and give its expectation form as preparation work before we prove the convergence of $V_0+\varepsilon V_1$.
\par Let $\hat{\sigma}_t$ be the worst case scenario volatility process and $\hat{X}_t^{\varepsilon}$ be the worst case scenario risky asset process. Then equations (\ref{e2.8}) can be rewritten as follows.
\begin{eqnarray}\label{e3.3}
\left\{ \begin{array}{rcl}
d \hat{X}_t^{\varepsilon}&=&r\hat{X}_t^{\varepsilon}dt+\hat{\sigma}_t \hat{X}_t^{\varepsilon}d W_t,\\
d \hat{Y}_t^{\varepsilon}&=&\hat{X}_t^{\varepsilon}dt.
\end{array}
\right.
\end{eqnarray}
\par We can get the expression of $\hat{\sigma}$ by equations (\ref{e2.10}) and $\hat{\sigma}(\varepsilon)=\sigma_0+\varepsilon \hat{\gamma}$, where
\begin{eqnarray}\label{e3.1}
\hat{\gamma}(t,x,y;\varepsilon)=
\left\{ \begin{array}{rl}
1,& \partial^2_{xx}V^{\varepsilon}(t,x,y)\geq0,\\
0,& \partial^2_{xx}V^{\varepsilon}(t,x,y)<0.
\end{array}
\right.
\end{eqnarray}
\par Similarly, by solving equations (\ref{e2.12}) of $V_1$, we have the volatility process: $\bar{\sigma}(\varepsilon)=\sigma_0+\varepsilon\bar{\gamma}$,
where
\begin{eqnarray}\label{e3.2}
\bar{\gamma}(t,x,y)=
\left\{ \begin{array}{rl}
1,& \partial^2_{xx}V_0(t,x,y)\geq0,\\
0,& \partial^2_{xx}V_0(t,x,y)<0.
\end{array}
\right.
\end{eqnarray}
Here, we use the short notation $\hat{\gamma}_t$ and $\bar{\gamma}_t$ for $\hat{\gamma}(t,x,y;\varepsilon)$ and
$\bar{\gamma}(t,x,y)$.

\par Let $Z^{\varepsilon}=V^{\varepsilon}-(V_0+\varepsilon V_1)$. To estimate the error term $Z^\varepsilon$, we define the operator $L(\sigma)=\partial_t+rx\partial_x-r+\frac{1}{2}\sigma^2x^2\partial^2_{xx}+x\partial_y$. According to partial differential equations
(\ref{e2.9}), (\ref{e2.11}) and (\ref{e2.12}), we have that
\begin{eqnarray*}\label{e3.5}
\begin{split}
L(\hat{\sigma}_t)Z^{\varepsilon}&=L(\hat{\sigma}_t)(V^\varepsilon-(V_0+\varepsilon V_1))\\
&=0-L(\hat{\sigma}_t)(V_0+\varepsilon V_1)\\
&=-(L(\hat{\sigma}_t)-L(\sigma_0))V_0-L(\sigma_0)V_0-\varepsilon(L(\hat{\sigma}_t)-L(\sigma_0))V_1
-\varepsilon L(\sigma_0)V_1\\
&=\varepsilon(\bar{\gamma}_t-\hat{\gamma}_t)\sigma_0x^2\partial^2_{xx}V_0-(\varepsilon^2/2)((\hat{\gamma}_t)^2
x^2\partial^2_{xx}V_0+2\sigma_0\hat{\gamma}_t x^2\partial^2_{xx}V_1)\\
&-(\varepsilon^3/2)(\hat{\gamma}_t)^2x^2\partial^2_{xx}V_1\\
&=-f^{\varepsilon}(t,x,y),
\end{split}
\end{eqnarray*}
with the boundary condition $Z^{\varepsilon}(T)=V^{\varepsilon}(T)-V_0(T)-\varepsilon V_1(T)=0.$
\par We have the following expectation form of $Z^{\varepsilon}$ by Dynkin formula.
 \begin{eqnarray*}\label{e3.6}
 Z^{\varepsilon}&=&E_{txy}\left[\int^T_tf^{\varepsilon}(s,x,y)ds\right]\\
 &=&\varepsilon E_{txy}\left[\int^T_t(\hat{\gamma}_s-\bar{\gamma}_s)\cdot\sigma_0\cdot(\hat{X}^{\varepsilon}_s)^2\partial^2_{xx}V_0(s,\hat{X}^{\varepsilon}_s,\hat{Y}^{\varepsilon}_s)d s\right]\\
 &~&+\varepsilon^2E_{txy}\left[\int^T_t\{\frac{1}{2}(\hat{\gamma}_s)^2(\hat{X}^{\varepsilon}_s)^2\partial_{xx}^2V_0(s,\hat{X}^{\varepsilon}_s,\hat{Y}^{\varepsilon}_s)\right.
 \\&~&+\sigma_0(\hat{\gamma}_s)(\hat{X}^{\varepsilon}_s)^2\partial^2_{xx}V_1(s,\hat{X}^{\varepsilon}_s,\hat{Y}^{\varepsilon}_s)\}d s{\bigg ]}\\
 &~&+\varepsilon^3E_{txy}\left[\int^T_t\frac{1}{2}(\hat{\gamma}_s)^2(\hat{X}^{\varepsilon}_s)^2\partial^2_{xx}V_1(s,\hat{X}^{\varepsilon}_s,\hat{Y}^{\varepsilon}_s)d s\right]\\
 &=&\varepsilon I_1+\varepsilon^2 I_2+\varepsilon^3 I_3,
 \end{eqnarray*}
 where
 \begin{eqnarray}
 I_1&=&E_{txy}\left[\int^T_t(\hat{\gamma}_s-\bar{\gamma}_s)\cdot\sigma_0\cdot(\hat{X}^{\varepsilon}_s)^2\partial^2_{xx}V_0(s,\hat{X}^{\varepsilon}_s,\hat{Y}^{\varepsilon}_s)d s\right], \label{e3.7}\\
 I_2&=&E_{txy}\left[\int^T_t\{\frac{1}{2}(\hat{\gamma}_s)^2(\hat{X}^{\varepsilon}_s)^2\partial_{xx}^2V_0(s,\hat{X}^{\varepsilon}_s,\hat{Y}^{\varepsilon}_s)\right.\nonumber
 \\&~&+\sigma_0(\hat{\gamma}_s)(\hat{X}^{\varepsilon}_s)^2\partial^2_{xx}V_1(s,\hat{X}^{\varepsilon}_s,\hat{Y}^{\varepsilon}_s)\}d s{\bigg ]},\label{e3.8}\\
 I_3&=&E_{txy}\left[\int^T_t\frac{1}{2}(\hat{\gamma}_s)^2(\hat{X}^{\varepsilon}_s)^2\partial^2_{xx}V_1(s,\hat{X}^{\varepsilon}_s,\hat{Y}^{\varepsilon}_s)d s\right].\label{e3.9}
 \end{eqnarray}
Thus we have that
\begin{eqnarray}\label{e3.10}
|Z^{\varepsilon}|\leq\varepsilon|I_1|+\varepsilon^2|I_2|+\varepsilon^3|I_3|.
\end{eqnarray}
So we can estimate $Z^{\varepsilon}$ by controlling $|I_1|$, $|I_2|$ and $|I_3|$.

\subsection{The polynomial growth condition of payoff function}
From section 4.2, we know that to control the error term, we need to analyze the three parts. By (\ref{e3.10}), we have
\begin{eqnarray*}
\left|\frac{Z^{\varepsilon}}{\varepsilon}\right|\leq|I_1|+\varepsilon(|I_2|+\varepsilon|I_3|).
\end{eqnarray*}
\quad Therefore, it is sufficient to prove
$$
\lim_{\varepsilon\downarrow0}|I_1|+\varepsilon(|I_2|+\varepsilon|I_3|)=0.
$$
\quad Obviously, it is necessary to give controls of $|I_2|$ and $|I_3|$. When it comes to $|I_1|$, we need to prove the convergence of it.
Now, let us consider controlling $|I_2|$ and $|I_3|$ first.
\par
By the expressions of $I_2$ and $I_3$, we can see that partial derivatives of $V_0$ and $V_1$ are involved. Thus, we should consider to estimate them before giving the controls of $I_2$ and $I_3$.
\par Next, we can obtain the expectation form of $V_0$ and $V^{\varepsilon}$ by the classical results.
When $\varepsilon=0$, we have
$$X(u)=x\exp\{(r-\frac{\sigma_0^2}{2})(u-t)+\sigma_0(W_u-W_t)\}.$$
Thus
\begin{eqnarray}\label{e3.11}
\begin{split}
V_0(t,x,y)&=e^{-r(T-t)}E_{txy}[\varphi(Y_{0,T})]\\
&=e^{-r(T-t)}E_{txy}\left[\varphi\left(\frac{1}{T}\int^T_0 X(u)du\right)\right]\\
&=e^{-r(T-t)}E_{txy}\left[\varphi\left(\frac{1}{T}\cdot x\cdot(\int^T_0e^{(r-\frac{\sigma_0^2}{2})(u-t)+\sigma_0(W_u-W_t)}du)\right)\right]\\
&=e^{-r(T-t)}E_{txy}[\varphi(x\cdot H)],
\end{split}
\end{eqnarray}
where $H(=(1/T)\int^T_0\exp\{(r-\sigma_0^2/2)(u-t)+\sigma_0(W_u-W_t)\}du)$ is a random variable for fixed $t\in[0,T]$.
\par Similarly, there is
\begin{eqnarray}\label{e3.12}
\begin{split}
V^{\varepsilon}(t,x,y)&=~e^{-r(T-t)}\underset{\sigma\in\mathcal{A}^{\varepsilon}}{\esssup}~\left\{E_{txy}\left[\varphi(Y_{0,T}^{\varepsilon})\right]\right\}\\
&=e^{-r(T-t)}E_{txy}\left[\varphi(x\cdot G)\right],
\end{split}
\end{eqnarray}
where $G(=(1/T)\int^T_0\exp\{(r-(\hat{\sigma}_{u})^2/2)(u-t)-\hat{\sigma}_{u}(W_u-W_t)\}du)$
is a random variable for fixed $t\in[0,T]$.
\par
By equation (\ref{e3.11}) and equation (\ref{e3.12}), we notice that it is necessary to impose polynomial growth conditions on $\varphi$ to control
$\partial^2_{xx}V_0$ and $\partial^2_{xx}V^{\varepsilon}$. Then we give the estimations of $\partial^2_{xx}V_0(t,x,y)$ and $\partial^2_{xx}V^{\varepsilon}(t,x,y)$ in following Lemma.

\begin{lemma}\label{4.1}
Suppose that the second derivative of payoff function satisfies the polynomial growth condition, i.e. there are constants $K_2$ and $m$ such that
$\varphi''(x)\leq K_2(1+|x|^m).$ Then, we have constant $K_3$ such that
\begin{eqnarray}\label{e3.13}
\left|\partial^2_{xx}V_0(t,x,y)\right|\leq K_3\left(1+|x|^m\right),
\end{eqnarray}
where $K_3$ depends on $T,~t,$ $E_{txy}\left[|H|^2\right]$, $E_{txy}\left[|H|^{m+2}\right]$ and $K_2$.
\par Moreover, there is constant $K_4$ such that
\begin{eqnarray}\label{e3.14}
\left|\partial^2_{xx}V^{\varepsilon}(t,x,y)\right|\leq K_4(1+|x|^m).
\end{eqnarray}
where $K_4$ depends on $T,~t,$ $E_{txy}\left[|G|^2\right]$, $E_{txy}\left[|G|^{m+2}\right]$ and $K_2$.
\end{lemma}
{\em Proof}.
As the assumption of $\varphi$ in the lemma, we have
\begin{eqnarray*}
\begin{split}
\left|\partial^2_{xx}V_0(t,x,y)\right|&=e^{-r(T-t)}E_{txy}\left[\varphi''(xH)H^2\right]\\
&\leq e^{-r(T-t)}E_{txy}\left[K_2(1+|xH|^m)H^2\right]\\
&\leq K_3\left(1+|x|^m\right).
\end{split}
\end{eqnarray*}
Here $K_3$ depends on $T,~t,$ $E_{txy}\left[|H|^2\right]$, $E_{txy}\left[|H|^{m+2}\right]$ and $K_2$.
\\Indeed, for a constant $m>0$, we have that
\begin{eqnarray*}
EH^m&=&\left(1/(T)\right)^mE\left(\int_{-t}^{T-t}\exp\{(r-\sigma_0^2/2)u+\sigma_0W_u\}du\right)^m\\
&\leq&(\frac{1}{T})^mE(\int_{-t}^{T-t}e^{|r-\sigma_0^2/2|(T-t)+\sigma_0W_u}du)^m\\
&\leq&(\frac{1}{T})^me^{m|r-\sigma_0^2/2|(T-t)}E\left(\sup_{s\in(-t,T-t)}\left\{e^{\sigma_0W_s}\right\}\right)^m<+\infty.
\end{eqnarray*}
\par On the other hand, we get the controls of $\partial^2_{xx}V^{\varepsilon}$ similarly.
Then there is a constant $K_4$ which depends on $T,~t,$ $E_{txy}\left[|G|^2\right]$, $E_{txy}\left[|G|^{m+2}\right]$ and $K_2$ such that
\begin{eqnarray*}
|\partial^2_{xx}V^{\varepsilon}(t,x,y)|\leq K_4(1+|x|^m).
\end{eqnarray*}
\hfill $\Box$

\par
Now, by following proposition, we can get the control of $I_2$ and $I_3$.
\begin{proposition}\label{i2i3}
 Assume that $\varphi\in\mathcal{C}_p^2(R^+)$ and satisfies Lipschitz continuity condition. Then there exist constants $C_1$ and $p_1$ such that $I_2, I_3$ in equation (\ref{e3.8}) and equations (\ref{e3.9}) satisfy
\begin{eqnarray*}
|I_2|+|I_3|\leq C_1(1+|x|^{p_1}).
\end{eqnarray*}
\end{proposition}
{\em Proof}.
By Lemma \ref{4.1}, we have the following inequality from (\ref{e2.10}) and (\ref{e3.14}).
\begin{eqnarray*}
|\partial_tV^{\varepsilon}+r(x\partial_xV^{\varepsilon}-V^{\varepsilon})+x\partial_yV^{\varepsilon}|&\leq&\left|\frac{1}{2}(\sigma_0+\varepsilon)^2x^2\partial^2_{xx}V^{\varepsilon}\right|\\
&\leq&\left|(K_4/2)(\sigma_0+\varepsilon)^2\left(|x|^2+|x|^{m+2}\right)\right|.
\end{eqnarray*}
By the expression of $V_1$, it is true that
\begin{eqnarray*}
|\partial_t V_1+r(x\partial_xV_1-V_1)+x\partial_y V_1|&\leq&\left|K_4\sigma_0\left(|x|^2+|x|^{m+2}\right)\right|.
\end{eqnarray*}
By (\ref{e2.12}) and (\ref{e3.13}), we get the controls of $x^2\partial^2_{xx}V_1$,
\begin{eqnarray}\label{e3.15}
\begin{split}
\left|x^2\partial^2_{xx}V_1\right|&=\left|\partial_tV_1+r(x\partial_xV_1-V_1)+x\partial_y V_1+\bar{g}_t\sigma_0x^2\partial^2_{xx}V_0\right|\cdot(2/\sigma_0^2)\\
&\leq\left(|\partial_t V_1+r(x\partial_xV_1-V_1)+x\partial_y V_1|+|\sigma_0x^2\partial^2_{xx}V_0|\right)\cdot(2/\sigma^2_0)\\
&\leq M_1\left(|x|^2+|x|^{m+2}\right),
\end{split}
\end{eqnarray}
where $M_1$ depends on $K_3$, $K_4$ and $\sigma_0$.
\par We can obtain the existence and uniqueness of $\hat{X}_t^{\varepsilon}$ from Theorem 5.2.1 in 
\cite{BO}.
Then, by the estimates of the moments of solutions of stochastic differential equations (Corollary 12 in Section 2.5 of 
\cite{NV}),
~there is a constant $N_1(q)$ for fixed $q>0$ such that
\begin{eqnarray}\label{e3.16}
E_{txy}\left[\sup_{s\in[t,T]}\left|\hat{X}_s^{\varepsilon}\right|^q\right]\leq N_1(q)e^{N_1(q)(T-t)}\left(1+|x|^q\right).
\end{eqnarray}
\par By (\ref{e3.9}), (\ref{e3.15}) and (\ref{e3.16}), we have the following inequality.
\begin{eqnarray}\label{e3.17}
\begin{split}
|I_3|&=\left|E_{txy}\left[\int^T_t\frac{1}{2}(\hat{\gamma}_s)^2(\hat{X}_s^{\varepsilon})^2\partial^2_{xx}V_1(s,\hat{X}_s^{\varepsilon},\hat{Y}_s^{\varepsilon})d s\right]\right|\\
&\leq (M_1/2)E_{txy}\left[\int^T_t(|\hat{X}_s^{\varepsilon}|^2+|\hat{X}_s^{\varepsilon}|^{m+2})d s\right]
\leq M_1'\left(1+|x|^{m+2}\right).
\end{split}
\end{eqnarray}
Here $M_1'$ depends on $T,~t$, $N_1(2)$, $N_1(m+2)$ and $M_1$.
\par By (\ref{e3.8}), (\ref{e3.13}), (\ref{e3.15}) and (\ref{e3.16}), we obtain the controls of $|I_2|$.
\begin{eqnarray}\label{e3.18}
\begin{split}
|I_2|&=\left|E_{txy}\left[\int^T_t\frac{1}{2}(\hat{\gamma}_s)^2(\hat{X}_s^{\varepsilon})^2\partial^2_{xx}V_0+\sigma_0(\hat{\gamma}_s)(\hat{X}_s^{\varepsilon})^2
\partial^2_{xx}V_1ds\right]\right|\\
&\leq(K_3/2)E_{txy}\left[\int^T_t(\hat{X}_s^{\varepsilon})^2+(\hat{X}_s^{\varepsilon})^{m+2}d s\right]\\
&~~~~~~+M_1E_{txy}\left[\int^T_t(\hat{X}_s^{\varepsilon})^2+(\hat{X}_s^{\varepsilon})^{m+2}d s\right]\\
&\leq M_2(1+|x|^{p_1}),
\end{split}
\end{eqnarray}
where $M_2$ depends on $T$, $t$, $M_1$, $K_3$, $N_1(2)$ and $N_1(m+2)$, $p_1\geq m+2$.
\par Combine (\ref{e3.17}) and (\ref{e3.18}), there is a constant $C_1$ such that
$$|I_2|+|I_3|\leq C_1(1+|x|^{p_1}).$$
\hfill $\Box$

\subsection{The continuity of the second derivative of payoff function}
By Proposition \ref{i2i3}, we obtain the controls of $I_2$ and $I_3$. Next, for fixed point $(t,x,y)\in[0,T]\times R^+\times R^+$, it suffices to prove that
$$\lim_{\varepsilon\downarrow0}|I_1|=0.$$
\par
Notice that, if $\varphi\in\mathcal{C}_p^2(R^+)$
(i.e. its derivatives up to order 2 have polynomial growth),
we can get following inequality by (\ref{e3.7}), (\ref{e3.13}),
 (\ref{e3.16}) and H\"older inequality,
\begin{eqnarray}\label{e3.181}
\begin{split}
|I_1|&\leq\left[E_{txy}\left[\int_t^T(\sigma_0(\hat{X}_s^{\varepsilon})^2\partial^2_{xx}V_0)^2ds\right]\right]^{1/2}
\left[E_{txy}\left[\int_t^T(\hat{\gamma}_s-\bar{\gamma}_s)^2ds\right]\right]^{1/2}\\
&\leq M_3(1+|x|^{p_2})^{1/2}\left[E_{txy}\left[\int_t^T|\hat{\gamma}_s-\bar{\gamma}_s|ds\right]\right]^{1/2}.
\end{split}
\end{eqnarray}
Here, $M_3$ depends on $K_3, T, t, \sigma_0$, and $p_2\geq4+2m$. Moreover, $M_3$ is independent of $\varepsilon$.
\par
Let $h^{\varepsilon}(t,x,y)=\hat{\gamma}(t,x,y;\varepsilon)-\bar{\gamma}(t,x,y)$. By (\ref{e3.1}) and (\ref{e3.2}), we have
\begin{eqnarray*}\label{e3.191}
|h^{\varepsilon}(t,x,y)|=\left\{ \begin{array}{rl}
1,&\partial^2_{xx}V^{\varepsilon}\partial^2_{xx}V_0<0,\\
0,&\partial^2_{xx}V^{\varepsilon}\partial^2_{xx}V_0\geq0.
\end{array}
\right.
\end{eqnarray*}
Thus, to prove $|I_1|\rightarrow0$ as $\varepsilon\rightarrow 0$, it suffices to prove that
\begin{eqnarray}\label{e3.192}
\lim_{\varepsilon\downarrow 0}E_{txy}\left[\int_t^T|h^{\varepsilon}(s,\hat{X}_s^{\varepsilon},\hat{Y}_s^{\varepsilon})|ds\right]=0.
\end{eqnarray}

\par By the expression of $h^{\varepsilon}$, we should analyze the derivatives of $V_0$ and $V^{\varepsilon}$. Here, we find that the continuity of $\varphi''$ is necessary.

\begin{lemma}\label{A11}
Assume that $\varphi''$ is continuous. Then $\partial^2_{xx}V_0$ and $\partial^2_{xx}V^{\varepsilon}$ are continuous with respect to $(x,y)$.
\end{lemma}
{\em Proof}.
By (\ref{e3.11}), we have $V_0(t,x,y)=e^{-r(T-t)}E_{txy}[\varphi(xH)]$ and $\partial^2_{xx}V_0(t,x,y)=e^{-r(T-t)}E_{txy}[\varphi''(xH)H^2]$.
If $\varphi''$ is continuous, then for all $x_0\in R^+$, $\delta>0$, there is a constant $\xi=\xi(\delta,x_0)$ such that
$$
|\varphi''(xH)-\varphi''(x_0H)|\leq\delta
$$
for all $xH\in(x_0H-\xi,x_0H+\xi)$.
\par So for all $(x_0,y_0)\in R^+\times R^+$, $xH\in(x_0H-\xi,x_0H+\xi)$, $y\in(y_0-\xi,y_0+\xi)$, we have
\begin{eqnarray*}
\begin{split}
|\partial^2_{xx}V_0(t,x,y)-\partial^2_{xx}V_0(t,x_0,y_0)|&=e^{-r(T-t)}|E_{txy}[\varphi''(x H)H^2-\varphi''(x_0H)H^2]|\\
&\leq e^{-r(T-t)}E_{txy}[H^2|\varphi''(x H)-\varphi''(x_0H)|]\\
&\leq e^{-r(T-t)}\delta E_{txy}[H^2].
\end{split}
\end{eqnarray*}
Thus we obtain $$\lim_{(x,y)\rightarrow(x_0,y_0)}\partial^2_{xx}V_0(t,x,y)=\partial^2_{xx}V_0(t,x_0,y_0).$$
\par Similarly, we can get the continuity of $\partial^2_{xx}V^{\varepsilon}$.
\hfill $\Box$
\begin{remark}
It is rational to think that $V^{\varepsilon}$ and its derivatives converge to $V_0$ and its corresponding derivatives as $\varepsilon$ approaching 0 by Lemma \ref{L}.
\end{remark}
\begin{remark}
To simplify the complexity brought by the variable $Y$, which is called path-dependence and to study the behavior of $h^{\varepsilon}$,
we define
$$D_{ty}^{\lambda}=\{x\in R^+~|~\partial^2_{xx}V^{\varepsilon_0}\partial^2_{xx}V_0\leq 0,~\exists \varepsilon_0>\lambda\}.$$ Let $D_{ty}^0=\lim\limits_{\lambda\downarrow 0}D_{ty}^{\lambda}$. Then we can get following equation when $\partial_{xx}^{2}V^{\varepsilon}$ is continuous,
\begin{eqnarray*}\label{e3.193}
D_{ty}^0=\{x\in R^+~|~\partial^2_{xx}V_0(t,x,y)=0\}.
\end{eqnarray*}
\end{remark}
\begin{remark}
To control $h^{\varepsilon}$, we divide $D_{ty}^{\lambda}$ into two parts. Let $\alpha(\rho)=[-\rho,\rho]$, we will disscuss the characters of $D_{ty}^{\lambda}\cap\alpha(\rho)$ and $D_{ty}^{\lambda}\cap\alpha(\rho)^c$.
\end{remark}

\begin{lemma}\label{A12}
Assume that $\varphi''$ is continuous. Then we have
$$P_{txy}(D_{ty}^0\cap \alpha(\rho))=0.$$
Here, $P_{txy}(\cdot)$ means the conditional probability taken with respect to $X_t^{\varepsilon}=x$, $Y_t^{\varepsilon}=y$.
\end{lemma}
{\em Proof}.
 By (\ref{e3.3}) and (\ref{e2.11}), we can get following equation
\begin{eqnarray}\label{e3.231}
\left\{ \begin{array}{rcl}
2\partial_t V_0+r(x\partial_x V_0-V_0)+\frac{1}{2}\sigma_0^2x^2\partial^2_{xx}V_0&=&0,\\
V_0(T)&=&\varphi(x H).
\end{array}
\right.
\end{eqnarray}
\par Let $Q=\partial^2_{xx}V_0$. Then by (\ref{e3.231}) we have
\begin{eqnarray*}\label{e3.241}
\left\{ \begin{array}{rcl}
2\partial_t Q+(r+\sigma_0^2)Q+(r+2\sigma_0^2)x\partial_x Q+\frac{1}{2}\sigma^2_0x^2\partial^2_{xx}Q&=&0,\\
Q(T)&=&\varphi''(x H)H^2.
\end{array}
\right.
\end{eqnarray*}
\par Let $x=\log k$. Then we have that
\begin{eqnarray}\label{e3.251}
\left\{ \begin{array}{rcl}
2\partial_t Q+(r+\sigma^2_0)Q+(r+2\sigma^2_0)\partial_{k}Q+\frac{1}{2}\sigma^2_0\partial^2_k Q&=&0,\\
Q(T)&=&\varphi''((\log k) H)H^2.
\end{array}
\right.
\end{eqnarray}
\par Notice that the coefficients in equations (\ref{e3.251}) are constants and Q is bounded on $D_{ty}^0\cap \alpha(\rho)$ by the continuity of $\varphi''$ and Lemma \ref{A11}. Moreover, by equations (\ref{e3.251}), we find that $y$ is not related to the equations. Then by Theorem A of \cite{SA}
and the remark below it,
we have that the number of zero points of $Q$ is

no more than countable for all $(s,y)\in[t,T]\times R$.

Thus, $\partial^2_{xx}V_0$ has no more than countable zero points.
\par Hence we have $P_{txy}(D_{ty}^0\cap \alpha(\rho))=0$ by Lemma 4.10 of \cite{Jean}.
Then the proof of Lemma \ref{A12} is completed.
\hfill $\Box$
\par On the basis of previous analysis, we will prove $(\ref{e3.192})$ now. We're going to split the expectation into two parts.
By proving the convergence of each part, we can get the convergence of the expectation.

\begin{proposition}\label{A31}
Assume that $\varphi\in\mathcal{C}_p^2(R^+)$ and $\varphi''$ is continuous. Then we obtain equation (\ref{e3.192}).
\end{proposition}
{\em Proof}.
Let $\bar{D}_{ty}^{\lambda}$ is closure of $D^{\lambda}_{ty}$,
$\bar{D}^0_{ty}=\lim\limits_{\lambda\downarrow 0}\bar{D}_{ty}^{\lambda}$ and $0\leq\lambda<\varepsilon<1$.
\par
By the definition of $D^{\lambda}_{ty}$, we have
\begin{eqnarray}\label{e3.261}
\begin{split}
&~~~~~E_{txy}\left[\int_t^T|h^{\varepsilon}(s,\hat{X}_s^{\varepsilon},\hat{Y}_s^{\varepsilon})|ds\right]\\
&\leq E_{txy}\left[\int^T_t\mathbb{I}_{\bar{D}^{\lambda}_{(s\hat{Y}^{\varepsilon}_s)}}(\hat{X}_s^{\varepsilon})ds\right]\\
&=E_{txy}\left[\int^T_t\mathbb{I}_{\bar{D}^{\lambda}_{(s\hat{Y}^{\varepsilon}_s)}\cap \alpha(\rho)}(\hat{X}_s^{\varepsilon})ds\right]+
E_{txy}\left[\int^T_t\mathbb{I}_{\bar{D}^{\lambda}_{(s\hat{Y}^{\varepsilon}_s)}\cap
\alpha(\rho)^c}(\hat{X}_s^{\varepsilon})ds\right]\\
&=\Phi_1+\Phi_2.
\end{split}
\end{eqnarray}
\par
Now, we consider the second part of (\ref{e3.261}) first. By (\ref{e3.16}) and Chebyshev's inequality, there is
\begin{eqnarray*}\label{e3.262}
\begin{split}
\Phi_2&\leq E_{txy}\left[\int^T_t\mathbb{I}_{\alpha(\rho)^c}(\hat{X}_s^{\varepsilon})ds\right]\\
&\leq \int^T_t P_{txy}\left(\sup_{s\in[t,T]}|\hat{X}_s^{\varepsilon}|\geq\rho\right)ds\\
&\leq((T-t)/\rho)E_{txy}\left[\sup_{s\in[t,T]}|\hat{X}_s^{\varepsilon}|\right]\\
&\leq\frac{(T-t)N_1(1)}{\rho}e^{N_1(1)(T-t)}(1+|x|).
\end{split}
\end{eqnarray*}
Thus, we have
\begin{eqnarray}\label{e3.263}
\lim_{\rho\rightarrow\infty}\Phi_2=0.
\end{eqnarray}
\par
When it comes to the first part, we note that
$$
\Phi_1=\int^T_tP_{txy}\left(\hat{X}^{\varepsilon}_s\in\bar{D}^{\lambda}_{(s\hat{Y}^{\varepsilon}_s)}\cap \alpha(\rho)\right)ds.
$$
\par
Let $\theta(\Omega)=\sup\limits_{\lambda\in[0,1]}P_{txy}(\Omega)$, then there is
\begin{eqnarray}\label{e3.271}
P_{txy}\left(\hat{X}^{\varepsilon}_s\in\bar{D}^{\lambda}_{(s\hat{Y}^{\varepsilon}_s)}\cap \alpha(\rho)\right)
\leq\theta\left(\bar{D}^{\lambda}_{(s\hat{Y}^{\varepsilon}_s)}\cap \alpha(\rho)\right).
\end{eqnarray}
\par
Notice that $\lambda<\varepsilon$. Then $\bar{D}^{\lambda}_{sy}$ is a sequence of decreasing closed sets as $\varepsilon\downarrow 0$. Obviously, $\hat{X}_s^{\varepsilon}$ converges weakly to $X_s$. Thus \{$X_s$\} is a weakly compact. By the Lemma 8
of 
\cite{LMS},
it can be seen that $\theta\left(\bar{D}^{\lambda}_{(s\hat{Y}^{\varepsilon}_s)}\cap\alpha(\rho)\right)
\downarrow\theta\left(\bar{D}^{0}_{(s\hat{Y}^{\varepsilon}_s)}\cap\alpha(\rho)\right)$ as $\varepsilon\downarrow 0$.
\par
By Lemma \ref{A11}, we have $\bar{D}^{0}_{sy}=D^0_{sy}$. Hence, there is
$$P_{txy}\left(\hat{X}^{\varepsilon}_s\in\bar{D}^{0}_{(s\hat{Y}^{\varepsilon}_s)}\cap\alpha(\rho)\right)=0.$$
Then by definition of $\theta(\Omega)$, we have
$$\lim_{\varepsilon\downarrow 0}\theta\left(\bar{D}^{\lambda}_{(s\hat{Y}^{\varepsilon}_s)}\cap\alpha(\rho)\right)=
\theta\left(\bar{D}^{0}_{(s\hat{Y}^{\varepsilon}_s)}\cap\alpha(\rho)\right)=0.$$
Thus, there is
$$\lim_{\varepsilon\downarrow 0}P_{txy}\left(\hat{X}^{\varepsilon}_s\in\bar{D}^{\lambda}_{(s\hat{Y}^{\varepsilon}_s)}\cap \alpha(\rho)\right)=0.$$
Then, we obtain
\begin{eqnarray}\label{e3.272}
\lim_{\varepsilon\downarrow 0}\Phi_1=0.
\end{eqnarray}
\par
By equation (\ref{e3.263}) and equation (\ref{e3.272}), for any $\delta>0$, there is  $\rho_0=\rho_0(t,x,y,\delta)>0$ such that
$$\Phi_2<\delta/2,~for~all~\rho>\rho_0.$$
Next, for given $\rho_0$ and $\delta$, there is $\varepsilon_0=\varepsilon_0(t,x,y,\delta,\rho_0(t,x,y,\delta))$ such that
$$\Phi_1<\delta/2,~for~all~\varepsilon<\varepsilon_0.$$
Therefore, for any $\delta>0$, there is $\varepsilon_0=\varepsilon_0(t,x,y,\delta)$ such that
$$\Phi_1+\Phi_2<\delta,~for~all~\varepsilon<\varepsilon_0,$$
i.e.
$$\lim_{\varepsilon\downarrow0}
E_{txy}\left[\int_t^T|h^{\varepsilon}(s,\hat{X}_s^{\varepsilon},\hat{Y}_s^{\varepsilon})|ds\right]=0.$$
\hfill $\Box$
\subsection{The proof of Main result}
Now, as the analysis above, we can give the brief proof of theorem \ref{main}.
\\{\em The proof of main result}.
By inequality (\ref{e3.181}) and Proposition \ref{A31}, we have
\begin{eqnarray}\label{e3.273}
\lim_{\varepsilon\downarrow0}|I_1|=0.
\end{eqnarray}
By inequality (\ref{e3.10}), we have
\begin{eqnarray*}
\left|\frac{V^{\varepsilon}-(V_0+\varepsilon V_1)}{\varepsilon}\right|\leq|I_1|+\varepsilon(|I_2|+\varepsilon|I_3|).
\end{eqnarray*}
By Proposition \ref{i2i3} and equation (\ref{e3.273}), we obtain the Theorem.
\hfill $\Box$
\section{Conclusion}\label{sec6}
 In this paper, we analyze the behavior of Asian option prices in the worst case scenario. The model studied in this paper is an uncertain volatility model with volatility interval $[\sigma_0,\sigma_0+\varepsilon]$. As $\varepsilon$ close to 0,
the ambiguity of model vanishes. We can also see that the worst case scenario prices of Asian option converge to its Black-Scholes prices with constant volatility as the interval shrinks. And through the study, we get an approach of estimating the worst case scenario Asian option prices. At the same time, it means that we give an estimation method to
solve a fully nonlinear PDE (\ref{e2.9}) by imposing additional conditions on the boundary condition and cutting it into two Black-Scholes-like equations.



\end{document}